\documentstyle[aps,pra,psfig,epsf,epsfig]{revtex}
\begin{document}

\twocolumn
\draft
\title{
Optimization approach to entanglement distillation
}
\author{T. Opatrn\'{y}$^{1,2}$ and G. Kurizki$^{1}$
 }
\address{
$^1$ Department of Chemical Physics, Weizmann Institute of Science,
761~00 Rehovot, Israel \\
$^{2}$ Department of Theoretical
Physics, Palack\'{y} University, Svobody 26, 779~00 Olomouc, \v{C}esko
}
\date{\today}

\maketitle
\begin{abstract}

We put forward a method for optimized distillation of  partly entangled pairs
of qubits
into a smaller number of more entangled pairs by recurrent local unitary
operations and projections. Optimized distillation is achieved by minimization
of a cost function with up to 30 real parameters, which is chosen to be
sensitive to the fidelity and the projection probability at each step. We show
that in many cases this approach can significantly improve the distillation
efficiency in comparison to the present methods.

\end{abstract}



\section{Introduction}

\begin{figure}[htb]
\centerline{ 
\begin{tabular}{cc}
\psfig{file=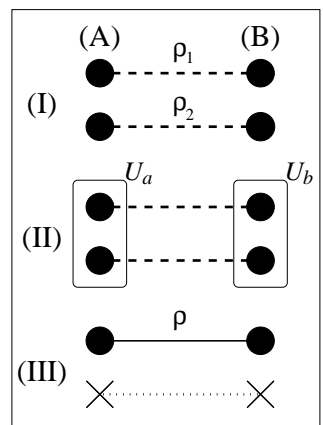,height=2.2in} 
\end{tabular}
}
\caption{
Stages of the basic entanglement purification scheme: 
(I) The two parties A and B share two
partly entangled pairs $\varrho_{1}$ and $\varrho_{2}$, (II)
they apply local unitary
transforms $U_{a}$ and $U_{b}$, and, (III) 
project out one of the pairs, thus obtaining
the remaining pair in the state $\varrho$.
}
\label{fig1b}
\end{figure}

Entanglement distillation 
or purification \cite{BBPSSW96,BDSW96,DEJMPS96,M98,HHH96}
denotes  the extraction of strongly entangled pairs 
of qubits from a larger number of
weakly entangled pairs. 
The objective is to share strongly correlated qubits between distant parties
in order to allow
reliable quantum teleportation \cite{telep} or quantum cryptography
\cite{crypt}.
All methods of entanglement purification require
that the two parties  perform only local operations on
their systems and that
only classical information be exchanged between them,
without transferring any
additional qubits.
The possible local operations include
(Fig.~\ref{fig1b}): (i)  unitary
transformations whereby each party entangles 
the particles at their
disposal; 
(ii) non-unitary projections, whereby  
each party measures a portion of their
particles, thus projecting  
the rest of the system onto a new state. 
These projections are usually
followed by classical communication of the measurement results between the
parties. Another non-unitary operation is
filtering  \cite{HHH96}, whereby
a pair is with a certain probability either 
discarded or kept after each projection, the probability
being dependent on the state.

The principles of the known distillation schemes have been surveyed in
\cite{BDSW96}.
Here we focus on recurrence distillation
methods which use two pairs of qubits as input and
produce a single output pair: 
(a)
The quantum privacy amplification
(QPA) method \cite{DEJMPS96}  requires that the projection 
of the input pairs on any 
of the four  states of the Bell basis ($|\psi^{\pm}\rangle$ = $2^{-1/2}$
$(0,1,\pm 1,0)^{\dag}$ and $|\phi^{\pm}\rangle$ = $2^{-1/2}$
$(1,0,0,\pm 1)^{\dag}$)
be larger than 1/2. The QPA method uses a sequence of $\pi/2$ rotations, 
controlled-not operations and measurements.
(b)
In the method of Ref.~\cite{BBPSSW96} 
the two-qubit input state is supposed to have a projection 
 larger than 1/2 on
the singlet state  $|\psi^{-}\rangle$. 
This method uses a sequence of unilateral $\pi$
rotations of the qubits and a bilateral XOR
operation 
followed by a measurement.
The output is 
a pair whose projection on 
the singlet state is  larger
than that of the input pair.
When used recurrently, the QPA method \cite{DEJMPS96}
often converges faster 
towards a Bell state than the method of \cite{BBPSSW96} (for the
proof of convergence of the QPA see \cite{M98}).

\begin{figure}[htb]
\centerline{ 
\begin{tabular}{cc}
\psfig{file=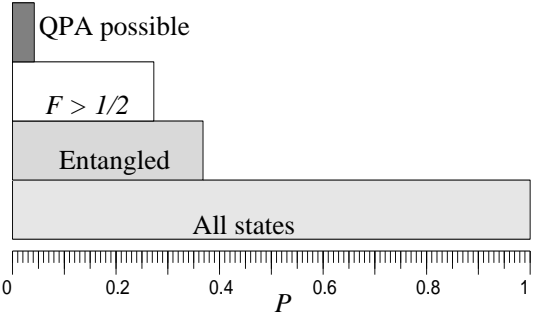,width=2.8in}
\end{tabular}
}
\caption{
Statistics of the entanglement properties of an
ensemble of randomly chosen two-qubit states.
The areas of the  
rectangles (from bottom to top)
are proportional to the 
following probabilities: the entire ensemble (All states), the
probability that a randomly
chosen state is entangled,
the probability that it fulfills the condition
$F$ $>$ $1/2$, and the probability that
it is directly usable as an input of the QPA scheme (QPA possible).
}
\label{fig1}
\end{figure}

Both of the aforementioned recurrence methods use {\em fixed}
parameters for the state transformations.
It is not 
clear whether these methods are
optimal for the states which are usable as their input, 
and, if not, how can their performance be improved.
There are other states for which  these methods do not work at all. 
This is demonstrated in Fig. \ref{fig1}, which gives
the statistics of applicability of the
QPA method 
for randomly chosen  two-qubit states. 
The random two-qubit density matrices were
generated as in Ref.~\cite{ZHSL98} and the statistics was obtained from a
sample of $10^{6}$ trials \cite{Slater98}.
For each of the generated density matrices 
it was checked whether it corresponds
to an entangled (i.e., inseparable)
state, and, if so, whether it can be purified by the QPA.
The results  show that about 37\% of 
all the possible states are 
 entangled and in principle can
be used for entanglement distillation. About 74\% of the entangled states
have fidelity 
(see Sec.~\ref{secqr}) \cite{BDSW96}  $F > 1/2$ 
and in principle can, after appropriate manipulations, be used for the QPA. 
Only a
 12\% fraction
 of the inseparable states, 
which is about 4\% of the sample, have a diagonal density-matrix 
element larger than 1/2  in the Bell basis,
and can therefore be used
directly as the input of the QPA algorithm.
This suggests that there is a vast domain in the space of two-qubit
states where new
approaches to the distillation problem can be useful.

Here we deal with  entanglement distillation as a problem of
optimization, aimed at improving the efficiency of the recurrence
methods and extending the class for which they work.
In contrast to the previously discussed recurrence methods, we assume that
the transformation parameters can be chosen at will to our advantage.
Furthermore, as opposed to the ingenious choice of parameters in these methods,
we do not have to guess their values:
the optimal choice of parameters for the unitary operations
and the projections is obtained by minimization of an approximate cost
function, which represents a tradeoff between maximized probability of the
conditional measurement and the best achievable entanglement. These
principles are similar to the ones previously used for optimized state
engineering \cite{costfunctions}. However, the present task is still
nontrivial, since the number of control parameters is large (up to 30), and
the choice of a cost function is far from obvious.

These issues are analyzed in Sec.~\ref{secopt}, where the optimization procedure
is obtained. This procedure is then applied in Sec.~\ref{secres} to cases where
the QPA is inefficient, to input states with fidelity $F$ $<$ 1/2, and to
trapped ion qubits. The results are discussed in Sec.~\ref{secdis}.


\section{Optimization procedure}
\label{secopt}

Let us assume that we want  to prepare, starting
from two pairs of qubits
with density matrices $\varrho_1$ and
$\varrho_2$, a single pair with a density matrix $\varrho$ whose
 entanglement is larger than that of both
$\varrho_1$ and $\varrho_2$,
using local unitary transformations followed by projections 
(Fig. \ref{fig1b}). 
Let us denote the local unitary transforms of each party 
by $U_{a}$ and
$U_{b}$, respectively. 
If the measurements are performed on the particles of the second pair, 
with the detected state being $|\Psi \rangle_{2}$, then
the first (purified) pair is transformed into the state
\begin{eqnarray}
  \label{opt0}
  \varrho = \frac{1}{P} ~_{2}\langle \Psi | 
  U_{a}  U_{b}\varrho_{1}
  \otimes \varrho_{2} U_{b}^{\dag}  U_{a}^{\dag} |\Psi \rangle _{2} ,
\end{eqnarray}
$P$ being the probability of success,  
\begin{eqnarray}
  \label{opt-prob}
  P = {\rm Tr} \ _{2}\langle \Psi | U_{a}  U_{b} \varrho_{1}
  \otimes \varrho_{2} U_{b}^{\dag}  U_{a}^{\dag} |\Psi \rangle _{2}.
\end{eqnarray}

Provided that
each particle is a two-level system, the two-particle unitary
transformation of each party belongs to the SU(4) group (considering SU(4)
instead of U(4) means that we omit the unimportant overall phase).
Such a group has 15 real parameters. The local transformations on both sides
are thus parametrized with 30 real parameters. 
Finding the optimum
values for these 30 parameters would enable us to perform the entanglement
distillation in the most efficient way.
For this purpose,
we must (i) find a proper parametrization of the transformations, (ii) choose a
function that quantifies the success of the distillation, and (iii)
have a suitable method for the optimization of this function.

\subsection{Parametrization of the local unitary transformations}

To parametrize the SU(4) local unitary transformations, we use a
modified version of the scheme in
Ref.~\cite{EP98}, in the form of a product of six SU(2) transformations:
\begin{eqnarray}
  \label{u1}
  U=&U^{(1,2)}&(\phi_{12},\psi_{12},\chi_{12})  \\
  \times &U^{(2,3)}&(\phi_{23},0,\chi_{23})
  U^{(1,3)}(\phi_{13},\psi_{13},\chi_{13}) \nonumber \\
  \times &U^{(3,4)}&(\phi_{34},0,\chi_{34})
  U^{(2,4)}(\phi_{24},0,\chi_{24}) \nonumber \\
  \times &U^{(1,4)}&(\phi_{14},\psi_{14},\chi_{14}) , \nonumber
\end{eqnarray}
where the $U^{(i,j)}(\phi_{ij},\psi_{ij},\chi_{ij})$ transforms represent
the SU(2) rotations between the $i$-th and $j$-th states of the
4-dimensional basis, their elements being 
\begin{eqnarray}
  \label{u1a}
  \begin{array}{ll}
   U^{(m,n)}_{k,k}=1 & k \neq m,n , \\
   U^{(m,n)}_{m,m}= \cos \phi e^{i \psi}, & 
   U^{(m,n)}_{m,n} = \sin \phi e^{i \chi}  \\
   U^{(m,n)}_{n,m}= -\sin \phi e^{-i \chi}, & 
   U^{(m,n)}_{n,n} = \cos \phi e^{-i \psi}.    
  \end{array}
\end{eqnarray}
We consider the basis states to be $|g_{1},g_{1}\rangle$,
$|g_{1},e_{2}\rangle$,
$|e_{1},g_{2}\rangle$, and $|e_{1},e_{2}\rangle$, where $|g_{k}\rangle$ and
$|e_{k}\rangle$ denote the ground and the excited state of the $k$th system,
$k=1,2$.

The local unitary transformations are thus
described by a 30-dimensional vector ${\bf X}$,
\begin{eqnarray}
  \label{u2}
  {\bf X} = \left( \phi^{a}_{12},\psi^{a}_{12},\chi^{a}_{12},
  \phi^{a}_{23},\chi^{a}_{23}, \right.
   \\
  \phi^{a}_{13},\psi^{a}_{13},\chi^{a}_{13},\phi^{a}_{34},\chi^{a}_{34},
  \phi^{a}_{24},\chi^{a}_{24},\phi^{a}_{14},\psi^{a}_{14},\chi^{a}_{14}; 
  \nonumber \\
  \phi^{b}_{12},\psi^{b}_{12},\chi^{b}_{12},\phi^{b}_{23},\chi^{b}_{23}, 
  \phi^{b}_{13},\psi^{b}_{13},\chi^{b}_{13},\phi^{b}_{34},\chi^{b}_{34},
  \nonumber \\
  \left.
  \phi^{b}_{24},\chi^{b}_{24},\phi^{b}_{14},\psi^{b}_{14},\chi^{b}_{14} 
  \right), \nonumber
\end{eqnarray}
where the indices $a$ and $b$ refer to the two parties (Alice and Bob,
respectively).
The distillation protocol of Ref~\cite{BBPSSW96}
corresponds to the vector
\begin{eqnarray}
  \label{u3}
  {\bf X}_{1}= \frac{\pi}{2}\left(1,0,0,0,0, 
  -1,0,0,0,0, 
  0,0,1,0,0;  \right. \\
  \left.  0,0,0,0,0,  
  1,0,0,0,0, 
 0,0,0,0,0\right)  \nonumber
\end{eqnarray}
(after the randomization yielding the Werner state), whereas
the QPA protocol 
\cite{DEJMPS96} is equivalent to the vector 
\begin{eqnarray}
  \label{u4}
  {\bf X}_{2}= \left( {\textstyle \frac{\pi}{4}},
  0,0,{\rm arcsin}{\textstyle \sqrt{\frac{2}{3}}},
  {\textstyle \frac{\pi}{2}}, \right. \\ 
  0,-{\textstyle \frac{\pi}{2}},0,{\textstyle \frac{\pi}{6}},
  {\textstyle \frac{\pi}{2}}, 
  {\rm arcsin}{\textstyle \sqrt{\frac{1}{3}}},{\textstyle \frac{\pi}{2}},
  {\textstyle \frac{\pi}{4}},0,0;  \nonumber \\
  {\textstyle \frac{\pi}{4}},0,0,{\rm arcsin}{\textstyle \sqrt{\frac{2}{3}}},
  -{\textstyle \frac{\pi}{2}}, \nonumber \\ 
  \left.  
  0,{\textstyle \frac{\pi}{2}},0,{\textstyle \frac{\pi}{6}},
  -{\textstyle \frac{\pi}{2}},
  {\rm arcsin}{\textstyle \sqrt{\frac{1}{3}}},
  -{\textstyle \frac{\pi}{2}},{\textstyle \frac{\pi}{4}},0,0 \right) . 
  \nonumber
\end{eqnarray}
Thus, both protocols represent specific choices of the available transformation
parameters.
The transformations with 15 parameters on each side are the most general
possible, but
even transformations with fewer degrees of freedom can be suitable for the
extremization of the distillation efficiency. 
The number of available parameters depends on the particular 
realization of the qubits and the way we manipulate them (see
Sec.~\ref{secions}).


\subsection{Quantifying the result of the distillation}
\label{secqr}

The resulting state should be as strongly entangled as possible and 
obtainable with a reasonably high probability. 
The calculation of the probability
of success in Eq. (\ref{opt-prob})
is easy, but quantifying the
entanglement is a non-trivial task 
for which many different measures 
have been suggested. 
Since finding the extremum
of a function is time consuming, we prefer a measure of entanglement
that can be calculated as fast as possible. 
This has led us
to choose as our measure the entanglement of
formation $E$ \cite{BDSW96}, defined as the least mean entanglement of ensembles
of pure states realizing the mixed state $\varrho$ (entanglement of the pure
state being the von Neumann entropy of the reduced  one-party density matrix)
 \cite{HW97}.
Another convenient measure of the entanglement is  fidelity
(or fully entangled fraction) $F$,
defined as the maximum ${\rm max} \langle e |\varrho | e\rangle$
taken over all completely entangled states $|e\rangle$ \cite{BDSW96,others}.

Along with the entanglement,
our ``cost'' function should optimize the probability of success.
We have experimented with the maximization of 
$E$ and $F$, and minimization of variously
constructed cost functions
depending on $F$, $E$ and the success probabilities $P$. 
The best results have been achieved by means of the cost
function
\begin{eqnarray}
   f_{c} = 1/[(\tilde F ^{q} \tilde P + \epsilon)(\tilde E + \epsilon)],
\end{eqnarray} 
where $\tilde F$ is the largest fidelity of the 4 possible outcomes of the
measurement, $\tilde P$ and $\tilde E$ are the corresponding probability and
entanglement, $\epsilon$ is a small parameter ($\approx 10^{-6}$) for
regularization of the function around $\tilde P$ $\approx 0$ and $\tilde E$
$\approx 0$, and $q$ is a parameter quantifying the preference for larger
fidelity or larger probability (typically, $q$ = 12).
The choice of this function ensures that a large entanglement is
achieved with a reasonable probability. By manipulating the shape of
the cost function 
(e.g., varying $q$)
we can get the resulting state with large entanglement
but small probability or vice versa with various
intermediate possibilities.

For a comparison of the results of different methods, it is useful to estimate
the average number of pairs which is consumed in order to get one pair with
the required entanglement. Assume first that we have reached the goal after 
$n_{k}$ steps. The joint probability of success in all steps is $P_{k}$,
which is the product of their individual success probabilities.
The index $k$ refers to the particular sequence of results in the 
individual steps (different sequences of intermediate states can lead to the
same required entanglement).
After each step (except the last one), 
the resulting state is taken as the input
state for the next step. 
If we have initially ${\cal N}$ pairs, then the
average
number of resulting pairs distilled following the sequence $k$ would be
\begin{eqnarray}
{\cal N}_{k} = \frac{P_{k}}{2^{n_{k}}} {\cal N} ,
\label{enka}
\end{eqnarray}
and the total average number of distilled pairs is
\begin{eqnarray}
{\cal N}_{\rm tot} = \sum_{k}  {\cal N}_{k},
\label{entot}
\end{eqnarray}
where the summation runs over all the sequences $k$ which lead to a pair with
the required entanglement.
The denominator $2^{n_{k}}$ in Eq. (\ref{enka}) reflects the fact that in each
step two pairs are consumed to produce one resulting pair. From equations
(\ref{enka}) and (\ref{entot}) it follows that the total number of pairs
needed to create one resulting pair is, on average
\begin{eqnarray}
N = \frac{\cal N}{{\cal N}_{\rm tot}} = \left( \sum_{k} 
\frac{P_{k}}{2^{n_{k}}}   \right) ^{-1} .
\end{eqnarray}


\subsection{Extremization procedure}

We have
searched for the extrema of the cost functions
numerically, using the Matlab procedure FMINS.
This procedure starts from a given point 
and 
uses the Nelder-Meade simplex search algorithm.
Since the extremized cost function is generally not
convex, the procedure finds a local extremum which need not be the global one.
Therefore, getting a result does not mean that we actually found the optimal
method for distillation. 
In our computations, we usually begin with several 
randomly generated starting points and choose the best result.
In general, we can call it a success if the
distillation efficiency exceeds that
of the methods suggested so far
\cite{BBPSSW96,DEJMPS96}.


\section{Results and applications}
\label{secres}

In order to compare our approach to the present methods
\cite{BBPSSW96,DEJMPS96}, we first applied the optimization procedure
to the class of states 
on which they mostly focus,
i.e., the Werner states \cite{Werner},
which are mixtures of the totally entangled and totally mixed states.
In this case the
numerical optimization brought no improvement and the QPA method seems to be
as efficient as ours for the Werner states.
By contrast, substantial improvement was found for states such that
the QPA method either converges
relatively slowly or cannot be used at all \cite{DEJMPS96}.
This refers to the states that do not have any
diagonal matrix
element that is larger than 1/2
in the Bell representation.
If one of the diagonal elements 
in  the Bell basis
is only slightly larger than zero,
the QPA convergence may be too slow for efficient applications.
Let us study these cases in more detail by 
considering characteristic examples.

\begin{figure}[htb]
\centerline{ 
\begin{tabular}{cc}
\psfig{file=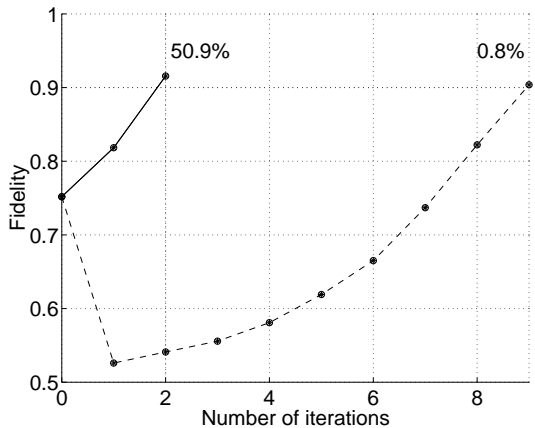,width=2.8in} 
\end{tabular}
}
\protect\caption{
Fidelity 
as a function of the number of iterations for the optimized scheme
(solid line) and for the QPA (dashed line)
with the initial state given by Eq.~(\ref{res1}).
Joint
probabilities of obtaining the 
required states are indicated. 
}
\label{fig-iter}
\end{figure}

\subsection{Cases when the QPA is inefficient}

As the first example, consider the state 
\begin{eqnarray}
  \label{res1}
  \varrho_{1} = \varrho_{2}
  = {\textstyle \frac{20}{33}}|\gamma\rangle \langle \gamma|+
 {\textstyle \frac{5}{33}}|\psi^{-}\rangle \langle \psi^{-}|+
 {\textstyle \frac{8}{33}\frac{1}{4}} I, 
\end{eqnarray}  
where 
$|\gamma \rangle$ = $2^{-1/2}(0,1,i,0)^{\dag}$ and $I$ is the 4$\times$4
identity operator.
The overlap of this state with the Bell state $|\psi^{-}\rangle$ is $17/33$
$\approx$ $0.5152$, which is 
marginally sufficient for using the QPA  algorithm, and the
fidelity defined above is $F$ = 0.7518. Let us assume that the
aim is to exceed  $F$ = 0.9, which may be sufficient
for application of other distillation schemes, e.g.,
the hashing method \cite{BDSW96}.
A direct application of the QPA method would reach this value after 9 steps,
the joint probability of success in all steps being 0.81\%
(see Fig.~\ref{fig-iter}). 
This would require
an average number of 63$\times 10^{3}$ pairs to get a single output pair.
On the other hand, our optimization scheme would reach the required fidelity in
2 steps with a joint probability of 50.87\%, so that less than 8 pairs on
average are consumed to get one output pair, which clearly
means a substantial improvement of distillation
efficiency \cite{webref}.


\subsection{Application to states with fidelity $<$ 1/2}

If the entangled pairs have fidelity $F$ $<$ 1/2, one cannot  
directly use the existing methods
\cite{BBPSSW96,DEJMPS96} to distill the entanglement.
So far, the only suggested scheme to handle such pairs would be
to transform them first by
non-unitary operations such as filtering \cite{HHH96},
in order to reach fidelity above
1/2. 
To our knowledge, no explicit formula for determining
the filtering parameters for arbitrary states has been presented. 

\begin{figure}[htb]
\centerline{ 
\begin{tabular}{cc}
\psfig{file=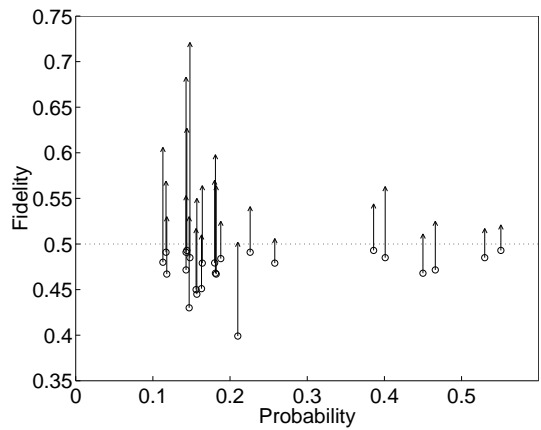,width=2.8in} 
\end{tabular}
}
\protect\caption{
Purification of inseparable states with $F$ $<$ 1/2. 
The vertical coordinate of the circle of each arrow denotes the fidelity of the
initial state, whereas that of its tip corresponds to the
fidelity of the resulting state after one step of optimized distillation. 
The horizontal coordinate of each arrow
indicates the probability of success of the distillation procedure.
The Figure shows 25 examples of the optimized
distillation procedure for randomly generated
initial states.
}
\label{fig3}
\end{figure}

Our approach allows for entanglement distillation 
of pairs with  $F$ $<$ 1/2
in the same way as for any other entangled states.
To demonstrate this,
we have randomly
generated several  density matrices with $E$ $>$ 0 and $F$ $<$ 1/2, and
optimized the local unitary transformations 
so as to reach a state with
$F$ $>$ 1/2 (see
Fig.\ref{fig3}).
We have observed that, whereas
the success of the optimization depends on the value of $E$ in the initial
state, 
it still works
 for all randomly generated
states with entanglement above 5$\times10^{-4}$,
allowing  
inseparable states with $F$ $<$ 1/2 to be purified using only unitary
transformations and conditional measurements, without filters.
Notice that starting from a fixed value of $F$, the distillation can exhibit
either a large increase of $F$ with a small probability or vice versa
\cite{webref}.


\subsection{Application to trapped ion qubits}
\label{secions}

Significant improvement in distillation efficiency is achieved by our method
not just for rather special, but also for generic,
physically important cases.
For instance, let us consider
qubits that are realized by two internal states of
trapped ions \cite{Cirac}. 
If two or more ions are trapped in a single trap, then the logical functions
between two qubits are achieved by using an auxiliary 
internal state of each ion 
and a vibrational mode of the collective oscillations.
It is assumed that the evolution of the ionic states is driven by
coherent laser pulses whose amplitudes, phases and durations can be controlled.
An arbitrary
unitary transformation of a single qubit can be achieved by two resonant
laser pulses
focused on the corresponding ion \cite{Cirac}. The durations (or strengths) 
of the two pulses and the phase of (say) the second pulse represent 3
parameters of the transformation. The interaction between two ions is 
achieved by a sequence of
three pulses, whose effect is  to flip the sign of the state 
$(0,0,0,1)^{\dag}$ (i.e., $| e_{1}, e_{2}\rangle$)
without changing the other basis states. 
The parameters of these three pulses are fixed at properly chosen values
so as to ensure that at the end of the procedure no auxiliary
state remains excited (see \cite{Cirac} for details). 

\begin{figure}[htb]
\centerline{ 
\begin{tabular}{cc}
\psfig{file=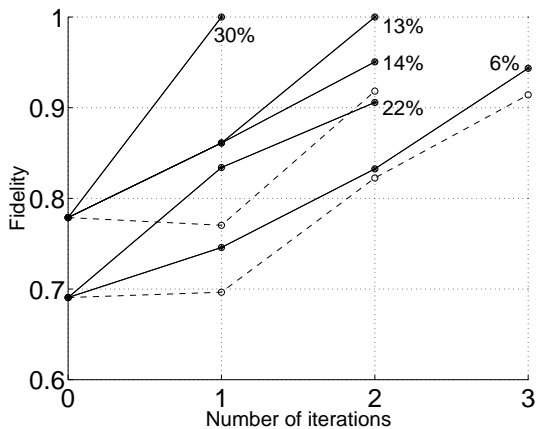,width=2.8in} 
\end{tabular}
}
\protect\caption{Fidelity 
as a function of the number of iterations for the optimized scheme
(solid lines) and for the QPA scheme
(dashed lines). Entanglement distillation for two
different initial states 
is repeated until the fidelity $F$ $=$ 0.9 is reached.
The upper tree (starting with $F$ = 0.779) corresponds 
to a decayed singlet (after a decay time $\gamma t$ $=$ 0.25)
as the initial state. The
lower tree (starting with $F$ = 0.691) corresponds to the 
same decayed singlet
mixed with a completely mixed state as the initial state. 
Joint probabilities of obtaining the 
required states are indicated for the optimized
scheme.
}
\label{fig2}
\end{figure}

Let us call this transformation $U_{c}$,
and assume that during one step of distillation each party
performs only one transformation $U_{c}$, preceded and followed by 
single-qubit
rotations. Rotation of all  4 qubits before the  $U_{c}$ transformation
represents 4 $\times$ 3 = 12 parameters. After the $U_{c}$ transformation
it is sufficient to rotate only the two qubits which are to be measured, as the
local transformations of the remaining pair do not change the entanglement.
Thus, we are left with 12 $+$ 2 $\times$ 3 = 18 parameters to be optimized, 
their
physical meaning  being the areas  and 
phases of the laser pulses.

As an example of entanglement decoherence, let us take the ionic
excitation to be
decaying according to the (zero-temperature) master equation
\begin{eqnarray}
  \label{master}
  \dot{\varrho} = -{\textstyle \frac{i}{\hbar}}
  \left[ H, \varrho \right] - {\textstyle \frac{\gamma}{2}} \left(
  \sigma_{+} \sigma_{-} \varrho + \varrho  \sigma_{+} \sigma_{-} \right)
  + \gamma \sigma_{-} \varrho \sigma_{+} .
\end{eqnarray}
Here the (single-particle)
Hamiltonian is $H$ $=$ $\hbar \omega_{0}$ $|e\rangle \langle e |$, with
$\sigma _{+}$ $=$ $|e\rangle \langle g |$
and $\sigma _{-}$ $=$ $|g\rangle \langle e |$, where $|g \rangle$ and
$|e\rangle$ denote the ground and the excited states, respectively.
The upper ``tree'' in Fig. \ref{fig2} has as its starting point the state
obtained by the decay of the singlet
$|\psi^{-}\rangle$ according to Eq. (\ref{master}), after a decay time
$\gamma t$ = 0.25. This state  has  fidelity
$F$ = 0.779.
Again, we can
assume that the aim is to purify the state so as to reach fidelity 
of at least
0.9. The QPA procedure would achieve this goal in 2 steps, consuming 
on average 7.7 pairs
and ending up with a state whose fidelity is 
$F$ = 0.918.
Using the optimization scheme with 18 parameters, 
the pairs
have a relatively large probability to be fully entangled 
($F$ $=$ 1.0) after the first or
second step. 
We are able to reach full entanglement in this case because
the decayed singlet 
has zero probability for both qubits to be in the excited state,
as opposed to a Werner state.
Following the different trajectories of our procedure, we find
that the mean number of pairs consumed before reaching $F$ $=$ 0.9 
is only  4.6.

The lower tree in Fig. \ref{fig2} starts from a state obtained by
mixing the decayed singlet  after $\gamma t$ $=$ 0.25
with a fully mixed state in a 83\% to 17\% proportion, as
a result
of additional sources of errors.  In this case, even the optimization scheme
yields only  partially entangled states. However,
whereas the QPA procedure
would need 25.8 pairs on average to get the required fidelity, the optimization
scheme would consume only 15.8 pairs on average  for this task \cite{webref}.


\section{Discussion}
\label{secdis}

Our main achievements can be characterized as finding a 
straightforward method
for efficient distillation of entanglement,
which is particularly valuable in cases where previously
suggested methods either do not work or converge relatively slowly. 
There is no special requirement on the form 
of the initial states (such as the Werner states), except that the states must
be entangled. It is not even required that the fidelity should be larger than
$1/2$. We have
seen that this approach allows essential saving of the ``raw material'' of
initial partly entangled states.

Several points must be noted: (i) The optimization 
procedure may end in a local extremum
of the cost function, which usually requires several trials before the final
choice of the transformation parameters. 
(ii) By contrast to the previous schemes \cite{BBPSSW96,DEJMPS96}, 
our method is state-dependent:
before starting the distillation we have to know the initial density matrix.
This knowledge is consistent with the objective of protecting particular
correlated two-qubit states, e.g., singlets, from being spoilt. The knowledge
of the initially spoilt state can be achieved either by familiarity with the
dissipation or error dynamics (e.g., zero-temperature decay -
Sec.~\ref{secions}), or by state reconstruction methods (the problem of
density matrix reconstruction by local measurements of  a pair of two-level
systems has been studied in detail in Ref.~\cite{vlado}). Of course,  in the
latter case a portion of the pairs will be consumed for the
reconstruction measurements, but
once the density matrix is determined with  sufficient precision, the
distillation scheme can run indefinitely. Note that the knowledge of the
density matrix would be necessary also in the case of the QPA method when $F$
$>$ $1/2$ but no diagonal density matrix element in the Bell
representation is larger than  $1/2$: the state must be properly rotated
before the QPA method itself is used, and the parameters of the
rotation would depend on the initial state.

\section*{Acknowledgments}

The authors are grateful to J.I.~Cirac, H.J.~Kimble,
T.~Mor, M.~Plenio, S.~Popescu, P.~Zoller, 
and K. \.{Z}yczkowski for discussions.
This work was supported by EU (TMR), ISF and Minerva grants.


\end{document}